\documentclass[12pt]{article}
\usepackage{amsfonts,amssymb,amsmath,enumerate}
\usepackage{theorem,cite}
\usepackage{indentfirst}
\usepackage{textcomp}
\usepackage{color}

\newtheorem{thm}{Theorem}[section]
\newtheorem{prop}{Proposition}[section]
\newtheorem{lemma}{Lemma}[section]
\newtheorem{coro}{Corollary}[section]

\theorembodyfont{\rmfamily}

\newcommand{\bea}{\begin{eqnarray}}
\newcommand{\eea}{\end{eqnarray}}
\newcommand{\beano}{\begin{eqnarray*}}
\newcommand{\eeano}{\end{eqnarray*}}
\newcommand{\beq}{\begin{equation}}
\newcommand{\eeq}{\end{equation}}
\newcommand{\nonu}{\nonumber \\}
\newcommand{\mb}[1]{\hspace{2.1ex}\mbox{#1}\hspace{2.1ex}}
\numberwithin{equation}{section}
\newcommand{\finproof}{{\hfill$\square$}}

\def\fs{{\mathfrak s}}
\def\ft{{\mathfrak t}}

        \def\cU{{\cal U}}

\newcommand{\CC}{{\mathbb C}}

\newcommand{\II}{{\mathbb I}}

\newcommand{\prt}{\partial}

\newcommand{\wt}[1]{\widetilde{#1}}

\newcommand\bra[1]{\left\langle #1\right|}
\newcommand\ket[1]{\left|#1\right\rangle}
\def\diag{\mathop{\rm diag}\nolimits}

\begin{document}
\pagestyle{empty}
\begin{flushright}
LAPTH-043/14
\end{flushright}

\vspace{20pt}

\begin{center}
\begin{LARGE}
{\bf $R$ matrices of three-state Hamiltonians\\[1.2ex] solvable by Coordinate Bethe Ansatz}
\end{LARGE}

\vspace{50pt}

\begin{large}
{T. Fonseca${}^{a}$, L. Frappat${}^a$, E. Ragoucy${}^a$ \footnote[1]{tiago.dinis@lapth.cnrs.fr, luc.frappat@lapth.cnrs.fr, eric.ragoucy@lapth.cnrs.fr}}
\end{large}

\vspace{15mm}

${}^a$  {\it Laboratoire de Physique Th\'eorique LAPTh, CNRS and Universit\'e de Savoie,\\
BP 110, 74941 Annecy-le-Vieux Cedex, France}

\end{center}

\vfill

\begin{abstract}
We review some of the strategies that can be implemented to infer an $R$-matrix from the knowledge of its Hamiltonian. 
We apply them to the classification achieved in \texttt{arXiv:1306.6303}, on three state $U(1)$-invariant Hamiltonians solvable by CBA, 
focusing on models for which the $S$-matrix is not trivial. 

For the 19-vertex solutions, we recover the $R$-matrices of the well-known Zamolodchikov--Fateev  and Izergin--Korepin models. We point out that the generalized Bariev Hamiltonian is related to both main and special branches studied by Martins in
\texttt{arXiv:1303.4010}, that we prove to generate the same Hamiltonian.  
The 19-vertex SpR model still resists to the analysis, although we are able to state some no-go theorems on its $R$-matrix.

For 17-vertex Hamiltonians, we  produce a new $R$-matrix. 
\end{abstract}

\vfill


\newpage
\pagestyle{plain}
\setcounter{page}{1}

\section{Introduction}

In his pioneering work \cite{Bethe}, Hans Bethe introduced the now called ``coordinate Bethe ansatz'' (CBA), allowing him to solve the Heisenberg model \cite{Heis}, i.e. to determine the eigenvalues and eigenvectors of the corresponding Hamiltonian. In the case of the one-dimensional quantum many-body problem with repulsive delta-function interaction, similar considerations led to the introduction of a new equation \cite{McGuire,Yang} that appears as a consistency condition for the factorization of the scattering matrix. It also showed up in Baxter's resolution of the eight vertex model \cite{Bax72}. This equation is now known as the Yang--Baxter equation (YBE) \cite{Yang,Baxt}.

A decade later, a crucial breakthrough arose from the works of the Leningrad school \cite{FST,FT}, see also \cite{KulSklya} and references therein. A new and more algebraic approach called quantum inverse scattering method (QISM) was developed, which constitutes nowadays the well-established framework for the study of quantum integrable systems. A particular object highlighted in this context is the quantum $R$-matrix \cite{D1,D2,J86}, satisfying the Yang--Baxter equation, see e.g. \cite{KuReSk,KuSkl}. Beyond the mathematical playground opened by the existence and properties of this key object, it has become the cornerstone of the study of quantum integrable systems. Indeed, in the context of one-dimensional spin chains, $R$-matrices with spectral parameters constitute the basic ingredient for constructing the monodromy matrix. It allows one also to define the transfer matrix and opens to the Algebraic Bethe Ansatz (ABA) \cite{FT}. The Yang--Baxter equation ensures the transfer matrices to commute for different values of the spectral parameter. The transfer matrix therefore encodes the conserved commuting quantities of the system, and among them the Hamiltonian, usually defined as the logarithmic derivative of the transfer matrix at a specific value of the spectral parameter. 

A Hamiltonian being given, it is therefore of great importance to determine whether there exists an $R$-matrix from which this Hamiltonian can be generated. 
Unfortunately, there is no systematic procedure to induce the form of the $R$-matrix once an integrable Hamiltonian is given. 
In this paper, we review some of the strategies that can be implemented to infer such solutions and we apply them to the classification achieved in \cite{r1d3}, on three state $U(1)$-invariant Hamiltonians solvable by CBA, focusing on models for which the $S$-matrix is not trivial. 
We also simplify the presentation of these Hamiltonians, relating the so-called telescoping terms and shifts by identity to Drinfeld twists leaving the Hopf structure unchanged.

For the four 19-vertex solutions, one recovers the $R$-matrices of the well-known Zamolodchikov--Fateev \cite{ZF} and Izergin--Korepin \cite{IK} models. They were solved through CBA and ABA in \cite{LS}. We point out that the third solution, the generalized Bariev Hamiltonian, is related to both main and special branches of \cite{M13}, which generate the same Hamiltonian.  
Finally, the 19-vertex SpR model still resists to the analysis, although we are able to state some no-go theorems for it.
For this latter model, we show that no univariate $R$-matrix can be associated to the corresponding Hamiltonian, while the question remains open for bivariate $R$-matrices due to the complexity of the calculations.
At this stage, it should be noticed that the equivalence between the CBA and QISM approaches, to the best of our  knowledge, has not been proved yet.

For 17-vertex Hamiltonians, we recover again the special branch $R$-matrix of \cite{M13}, but at a special point, and show how the 19-vertex solution degenerate to 17-vertex at this special point. We also produce a new $R$-matrix, associated to the Hamiltonian called $17V_2$ in \cite{r1d3}. It could be of some interest to study the physical content of this new Hamiltonian.

 There is also a 14-vertex solution with a non-trivial scattering matrix. Surprisingly enough, we prove that it does not exist any (univariate or bivariate) $R$-matrix for generic values of the parameters of the Hamiltonian. Hence, this case could be of some relevance in the comparison between CBA and ABA methods.

The paper is organized as follows. In section \ref{sect:setup}, we fix the main notations and properties of $R$-matrices. In section \ref{sect:constr}, we review some constructions of the $R$-matrices, namely the Baxterization procedure, the iteration procedure, the resolution by brute force and the spectral curve approach. We emphasize when relevant the behavior of the $R$-matrix whether it is univariate or bivariate. In section \ref{sect:resu}, we expose the results for the Hamiltonians under consideration.

\section{General set-up}
\label{sect:setup}

\subsection{Notations}

We consider $U(1)$-invariant Hamiltonians $H$ acting on a spin chain of length $L$ with nearest neighbour interactions and assuming periodic conditions for the chain (sites $L+1$ and 1 are identified), that is 
\beq
H=\sum_{j=1}^L H_{j,j+1}.
\label{eqH1}
\eeq
The $U(1)$ symmetry of $H$ is generated by the $S^z$ component of the total spin:
\beq
[H\,,\,S^z]=0\,,\quad S^z=\sum_{\ell=1}^L \fs^z_\ell
\mb{with} \fs^z|j\rangle =j\,|j\rangle.
\eeq
We are interested in models where the two-site Hamiltonian $H_{j,j+1}$ describes a three-state system, in other words $H_{j,j+1}$ acts in $\CC^3$ as a vector space, with basis vectors $|0\rangle = (1,0,0)$, $|1\rangle = (0,1,0)$, $|2\rangle = (0,0,1)$. If $E_{ij}$ denote the elementary $3 \times 3$ matrices with entry 1 in position $(i,j)$ and zero elsewhere, the two-site Hamiltonian under consideration reads
\begin{align}
H_{12} = {} & \sum_{i_1,i_2,j_1,j_2 \in \{0,1,2\}} h_{i_1\;i_2}^{j_1\;j_2} \, E_{i_1,j_1} \otimes E_{i_2,j_2} \nonumber \\
= {} & p E_{01} \otimes E_{10} + q E_{10} \otimes E_{01} + t_1 E_{21} \otimes E_{01} + s_1 E_{12} \otimes E_{10} + t_2 E_{01} \otimes E_{21} + s_2 E_{10} \otimes E_{12} \nonumber \\
&+ t_3 E_{12} \otimes E_{21} + s_3 E_{21} \otimes E_{12} + t_p E_{02} \otimes E_{20} + s_p E_{20} \otimes E_{02} + \sum_{i,j} v_{ij} E_{ii} \otimes E_{jj}\,.
\label{eqH2}
\end{align}

As already mentionned, Hamiltonians of the form \eqref{eqH1}-\eqref{eqH2} and 
solvable by CBA have been classified in \cite{r1d3}. The next question is then whether these Hamiltonians can be related to an $R$-matrix, using the transfer matrix formalism.
More precisely, let $\check R(x,y)$ be a (general bivariate) solution of the braided Yang--Baxter equation (YBE):
\begin{equation}\label{YBE}
\check R_{12}(y,z) \check R_{23}(x,z) \check R_{12}(x,y) = \check R_{23}(x,y) \check R_{12}(x,z) \check R_{23}(y,z) \,.
\end{equation}
We will assume that the $R$-matrix is unitary 
\beq\label{unitarite}
\check R(x,y) \check R(y,x) \propto \II,
\eeq
and regular
\beq\label{regularite}
\check R(x,x) = \II.
\eeq 
The transfer matrix associated to this $R$-matrix reads
\beq
t_{<12...L>}(x|y_1,...,y_L) = Tr_0 R_{01}(x,y_1)\,R_{02}(x,y_2)\,\cdots R_{0L}(x,y_L).
\eeq
The regularity and unitary conditions on $R$ ensure the existence of local interaction 
\beq
H=\left.\frac{d}{dx} \ln t_{<12...L>}(x|y,y,...,y)\right|_{x=y}=\sum_{j=1}^L H_{j,j+1},
\eeq
the local Hamiltonian being given by 
\begin{equation}\label{H=dR}
H_{12} = \left. \partial_x \check R_{12}(x,y) \right|_{x=y} \,.
\end{equation}

 In most cases, the $R$-matrix will be \emph{univariate}: \emph{multiplicative}, $\check R(x,y)=\check R(x/y)$, or \emph{additive}, $\check R(x,y)=\check R(x-y)$. However, we will also encounter cases where the $R$-matrix is genuinely \emph{bivariate}, see below.
We will also use the non braided $R$-matrix defined by $R = P \check R$ where $P$ is the permutation operator.
 
\subsection{Transformations on $R$-matrices and corresponding Hamiltonians}
\label{subsect:transfo}

Before computing the $R$-matrix, one can note that there exists transformations on $H$ that lead to physically equivalent models. 
We shall show that most of the transformations can be re-interpreted as some particular Drinfeld twists that preserve the Hopf structure (hence the Yang--Baxter equation). We remind the form of a Drinfeld twist (once represented):
\beq
\check R(x,y) \to \check R^F(x,y) = F(x,y)\, \check R(x,y)\, F^{-1}(y,x)
\mb{where} F(x,y)\in \text{End}(\CC^3\otimes \CC^3)[x,y]
\eeq
The following particular Drinfeld twists can be related to such transformations.

\paragraph{Factorized twist / gauge transformation.}

Let $F = g\otimes g$, $g\in \text{End}(\CC^3)$. It is known that the transformed $\check R^F$-matrix satisfies YBE. The two-site Hamiltonian is then transformed as 
\begin{equation}
H_{j,j+1} \to g \otimes g \; H_{j,j+1} \; g^{-1}\otimes g^{-1} \,.
\end{equation}

\paragraph{Grading twist / rescaling of parameters.}

Let $F = g \otimes g^{-1}$ with $g=\exp({\alpha \fs^z})$. It can be checked that since $[g\otimes g, \check R] = 0$, the  $\check R^F$-matrix satisfies  YBE.
The two-site Hamiltonian is then transformed as 
\begin{equation}
H_{j,j+1} \to e^{\alpha \fs^z} \otimes e^{-\alpha \fs^z} \; H_{j,j+1} \; e^{-\alpha \fs^z} \otimes e^{\alpha \fs^z} \,.
\label{eq:gradation}
\end{equation}

\paragraph{Conjugation / telescopic terms.}

Let $F(x,y) = g(x)\otimes g(y)$, where $g(x)=\exp(-xA)$ and $A$ is a diagonal matrix. It is easy to see that the  $\check R^F$-matrix satisfies YBE. The two-site Hamiltonian is then transformed as:
\begin{equation}
H_{j,j+1} \to H_{j,j+1} + A \otimes \II - \II \otimes A  \,,
\end{equation}
which generates a generic telescopic term.

 \paragraph{Normalization / shifts by identity.}
Rescaling the $\check R$-matrix by a function $f(x,y)=\exp(\alpha(x-y))$ amounts to shift the Hamiltonian as
\beq
H_{j,j+1} \to H_{j,j+1} + \alpha\,\II\otimes\II
\eeq

Finally, let us note that since $H$ and $S^z$ commute, one can also shift $H$ by $S^z$. 
However, this transformation is not related to a transformation on $R$ that preserves the Hopf structure.
Hence, in the course of reconstructing an $R$-matrix from $H$, one has to deal with $H+\beta\,S^z$ 
and tune the parameter $\beta$.

\section{Methods for constructing $R$-matrix}
\label{sect:constr}
There is no general constructive procedure to obtain an $R$-matrix from an integrable Hamiltonian. However, there are some techniques that may (or may not) work, depending on the considered Hamiltonian. We briefly review them, adding some properties for some of them.

\subsection{Baxterization}

The method of Baxterization has been proposed by V.F.R. Jones \cite{Jones}. It allows one to obtain solutions of the Yang--Baxter equation with spectral parameter from representations of the braid group, in particular in the Hecke, Temperly--Lieb and Birman--Murakami--Wenzl cases \cite{Isaev,KMN,Jim86}. Note that beyond this procedure, many authors tried to generalize or produce other suitable formulae that may lead to solutions of the YBE \cite{CGX,ZGB,YQL}.

Consider the braid group $\mathcal{B}_N$ generated by generators $T_i$ ($i=1,\ldots,N-1$), their inverses $T_i^{-1}$ and the relations (see~\cite{CP}):
\begin{equation}
\begin{split}
& T_i \, T_{i+1} \, T_i = T_{i+1} \, T_i \, T_{i+1} \;, \\
& T_i \, T_j = T_j \, T_i \quad \mbox{for} \ |i-j| > 1.
\end{split}
\label{BG}
\end{equation}
We set $Z_i=T_i-T_i^{-1}-\xi$ for some constant $\xi$.

\paragraph{Hecke case:} When the $T_i$'s satisfy the supplementary quadratic relations $Z_i = 0$, then 
\begin{equation}
\check R_{i,i+1}(z) = zT_i - z^{-1}T_i^{-1}
\end{equation}
is unitary and satisfies the Yang--Baxter equation with multiplicative spectral parameter $z$.

\paragraph{Temperly--Lieb case:} When $\ft_i=T_i+1$ satisfy the supplementary relations
\begin{equation}
 \ft_i \, \ft_{i\pm1} \, \ft_i = \ft_i \mb{and}
 \ft_i\, \ft_i  = 2a\,\ft_i,
\label{TL}
\end{equation}
for some parameter $a$, then 
\begin{equation}
\check R_{i,i+1}(z) = \ft_i -a +\frac{z+1}{z-1}\sqrt{a^2-1}
\end{equation}
is unitary and satisfies the Yang--Baxter equation with multiplicative spectral parameter $z$.

\paragraph{Birman--Murakami--Wenzl case:} When $Z_i$ and $T_i$ satisfy the following supplementary relations 
\begin{equation}
 Z_i \, T_{i-1}^{\pm 1} \, Z_i = \xi a^{\mp 1} Z_i \,, \quad Z_i \, T_{i}=T_i \, Z_i = - a\, Z_i 
 \end{equation}
for some constant $a$ (BMW algebra), both matrices 
\begin{equation}
\check R^{(\pm)}_{i,i+1}(z) =  T_i + \frac{a_\pm z^2}{1-a_\pm z^2}\, Z_i + z  \frac{q-1/q}{z-1/z}\, \II \mb{with} \xi = q-\frac1q
\mb{and} a_\pm= \mp aq^{\pm1}
\end{equation}
are unitary and satisfy the Yang--Baxter equation with spectral parameter $z$.

\bigskip

In all cases, eq. \eqref{H=dR} leads to $H=T_i$.
Hence the Baxterization procedure is the simplest way to get an $R$-matrix from an Hamiltonian, but it works only for specific Hamiltonians satisfying the braid group relations and Hecke, Temperly--Lieb or BMW algebra relations.

\subsection{Iteration procedure \textsl{\`a la Idzumi et al.}}

In this section we review Idzumi's method \cite{ITA} to construct solutions of the Yang--Baxter equation and derive some interesting properties that can help for simplifying the problem. Idzumi and collaborators used this method to build 19-vertex solutions, but for univariate $R$-matrices only. They found however four new Hamiltonians, that were solved in \cite{KMZ} through TQ relations.
Unfortunately the method is not so efficient for bivariate $R$-matrices. 
We will give below a generalization that works for the latter case, but some freedom is left that cannot be resolved without any further assumption on $R$. Moreover, in both cases (univariate or bivariate), the $R$-matrix is obtained as a series that may be difficult to handle.

\subsubsection{Recursion formulae for multiplicative $R$-matrix}
In what follows, we consider the case of $R$-matrices with multiplicative spectral parameters, although the original paper deals with additive ones. 

Let $\check R(u)$ be a solution of the multiplicative Yang--Baxter equation:
\begin{equation}
\label{eq:YBE_multiplicative}
\check R_{12} (u) \check R_{23} (uv) \check R_{12} (v) = 
\check R_{23} (v) \check R_{12} (uv) \check R_{23} (u) \,,
\end{equation}
and suppose that $\check R (u)$ is analytical around $u = 1$:
\begin{equation}
\check R (u) = \sum_{i=0}^\infty \check R^{(i)} (u-1)^i \,.
\end{equation}
We demand regularity for $R$ and thus $\check R^{(0)} = \II \otimes \II$.
Furthermore, the Hamiltonian being defined by the first derivative, one has $\check R^{(1)} = H$. 

\begin{thm}\cite{ITA}
Let $\check R(u)$ be an analytical solution of the multiplicative Yang--Baxter equation \eqref{eq:YBE_multiplicative}.
An Hamiltonian $H$ being given, the full matrix $\check R(u)$ such that $\check R^{(1)} = H$ can be reconstructed, up to an arbitrary normalization factor.
\label{thm:idzumi}
\end{thm}
\emph{Proof}.
We perform a Taylor expansion of \eqref{eq:YBE_multiplicative} and select the coefficient of $(u-1)^k (v-1)$, for $k \geq 1$.
Using $\check R^{(0)} = \II \otimes \II$, we get:
\begin{align}
(k+1) \left( \check R_{12}^{(k+1)} - \check R_{23}^{(k+1)}\right) 
= {}& \sum_{j=0}^k \left( \check R_{12}^{(j)} \check R_{23}^{(k-j)} H_{12} - H_{23} \check R_{12}^{(k-j)} \check R_{23}^{(j)} \right) \nonumber \\
& + \sum_{j=0}^k (k-j) \left( \check R_{12}^{(j)} \check R_{23}^{(k-j)} - \check R_{12}^{(k-j)} \check R_{23}^{(j)}\right) \nonumber \\
& + \sum_{j=1}^k (k-j+1) \left( \check R_{12}^{(j)} \check R_{23}^{(k-j+1)} - \check R_{12}^{(k-j+1)} \check R_{23}^{(j)}\right).
\end{align}
Only terms of order less than $k$ appear in the right hand side. Thus we write this equation as
\begin{equation}
\label{eq:rec_multiplicative}
\check R_{12}^{(k+1)} - \check R_{23}^{(k+1)} = Q^{(k)} \,,
\end{equation}
where $Q^{(k)}$ only depends on lower terms, so that the system is triangular, expressed on matrices in $\text{End}(\CC^3\otimes\CC^3\otimes\CC^3)$. 
Any solution of the Yang--Baxter equation~\eqref{eq:YBE_multiplicative} can be normalized such that $\check R_{aa}^{aa}(u) = 1$, 
that is $\check R_{aa}^{aa,(k+1)} = 0$ for all $k \in \mathbb N$, for some given $a$. This corresponds to the arbitrary 
normalization factor. 
Then, looking at the different entries, i.e. computing $\bra{abc} (\ldots) \ket{def}$, where $(\ldots)$ represents 
eq.~\eqref{eq:rec_multiplicative}, one can deduce the entries of the matrix at level $k+1$ from the ones at level $k$:
\bea
\check R_{ab}^{ab,(k+1)} &=& -\bra{aab} Q^{(k)} \ket{aab} \,,\quad \forall a,b
\\
\check R_{bd}^{ce,(k+1)} &=& - \bra{abd} Q^{(k)} \ket{ace} \,,\quad d \neq e
\\
\check R_{bd}^{ce,(k+1)} &=& \bra{bda} Q^{(k)} \ket{cea} \,,\quad b \neq c
\eea
\finproof

\subsubsection{Specific case of $U(1)$-invariant models}
Having established that the computation of the matrix $\check R$ can be achieved through recursion formulae in a rather general framework, we now restrict ourselves to the specific case of 19-vertex models. 
\begin{prop}
\label{prop:ice-type}
Let $H$ be a spin preserving Hamiltonian:
\[
[H, \fs^z \otimes \II + \II \otimes \fs^z] = 0 \mb{i.e.} H_{ab}^{cd} = 0 \quad \textrm{if} \quad a+b \neq c+d
\mbox{ (ice rule condition)}\,.
\]
If $H$ is obtained from a multiplicative $R$-matrix,  this $R$-matrix must also preserves the spin
\[
[ R(u), \fs^z \otimes \II + \II \otimes \fs^z] = 0 \,.
\]
\end{prop}

\emph{Proof}.
We proceed by induction. Suppose that the ice-rule property is satisfied by the matrix $\check R^{(k)}$ at order $k$ and use the relation \eqref{eq:rec_multiplicative}.
Computing an entry $\check R_{ab}^{cd,(k+1)}$ where $c+d \neq a+b$ amounts to compute 
\[
\check R_{bd}^{ce,(k+1)} = - \bra{abd} Q^{(k)} \ket{ace} 
\mb{and} \check R_{bd}^{ce,(k+1)} = - \bra{bda} Q^{(k)} \ket{cea} \,.
\]
In any case, the right hand side vanishes because $Q^{(k)}$ preserves the spin.
Moreover the property is obviously true for $k=1$, hence $\check R^{(k)}$ satisfies the ice-rule property for all $k$. Therefore the proposition is proved.
\finproof

So, if we start with an ice-type Hamiltonian, we only need to use the part of the recursion such that $a+b = c+d$, hence avoiding most of the computations.

One can wonder whether it is possible to go further in the simplification of the $R$-matrix when considering Hamiltonians that have more zero entries, i.e. if some zeros are preserved when computing $\check R(u)$ from $H$. One can show the following:
\begin{prop}
Let $H$ be a spin preserving Hamiltonian, that satisfies the additional constraint $H_{ce}^{bd} = 0$ for all $b \ne c$ for fixed $c$. Then 
\[
\check R_{ce}^{bd} (u)= 0 \qquad \textrm{for all }b \neq c \,.
\]
\end{prop} 

\emph{Proof}. Suppose that the property is true for $\check R^{(k)}$ at order $k$. We use the recursion relation \eqref{eq:rec_multiplicative} to compute the next term:
\[
\check R_{ce}^{bd,(k+1)} = \bra{bda} Q^{(k)} \ket{cea} \,.
\]
The right hand side is composed by terms like 
$\bra{bda} \check R_{12}^{(j)} \check R_{23}^{(i)} H_{12} \ket{cea}$, 
$\bra{bda} H_{23} \check R_{12}^{(i)} \check R_{23}^{(j)} \ket{cea}$, 
$\bra{bda} \check R_{12}^{(j)} \check R_{23}^{(i)} \ket{cea}$,
where $i,j \leq k$. All of them vanish trivially.
The property being obviously true for $k=1$, the result is proved.
\finproof

This property is valid for any entry, for instance
\begin{align*}
\textrm{For }b \textrm{ fixed}, \quad H_{ce}^{bd} = 0 \quad \textrm{for all }c \neq b \quad \Rightarrow \quad \check R_{ce}^{bd} (u) = 0 \quad  \textrm{for all }c \neq b  \,.
\end{align*}
In particular, one can deduce:
\begin{coro}
 The multiplicative matrices $\check R(u)$ that may lead to the 14-vertex Hamiltonians of Ref. \cite{r1d3}  can have at most $15$ non-zero entries.
\end{coro}

\subsubsection{Recursion formulae for bivariate $R$ matrices}

We focus now on a more general case by considering bivariate $R$ matrices, i.e. that depend on two spectral parameters.
We start by reviewing some basic results. 
\begin{lemma}
\label{prop:RR=Id}
Let $\check R(x,y)$ be a solution of the braided Yang--Baxter equation \eqref{YBE}, that is regular. Then $\check R(x,y)$ is unitary:
\beq\label{unit:general}
\check R(x,y) \check R(y,x) = \lambda(x,y) \,\II \otimes \II \,,
\eeq
where $\lambda$ is some symmetric scalar function. 
\end{lemma} 
\emph{Proof}. Equation \eqref{YBE} taken at $z=x$ gives
\begin{equation}
\label{eq:pre_identity}
\check R_{12} (y,x)\check R_{12} (x,y) =
\check R_{23} (x,y)\check R_{23} (y,x) \,.
\end{equation}
that is $M_{12} (y,x) \otimes \II_3 = \II_1 \otimes M_{23} (x,y)$ with $M(x,y) = \check R (x,y) \check R (y,x)$. 
This equality shows that $M_{23}(x,y)$ is proportional to identity in space 3, while $M_{12}(y,x)$ is proportional to identity in space 1. This implies that $M(x,y)=\lambda(x,y)\,\II\otimes\II$, where $\lambda(x,y)$ is some scalar function that must be symmetric.
\finproof
\begin{lemma}
If $\check R(x,y)$ is a regular solution of the braided Yang--Baxter equation \eqref{YBE}, then the two Hamiltonians
\beq
H=\prt_x \check R(x,y)\Big|_{x=y} \mb{and} \wt H=\prt_y \check R(x,y)\Big|_{x=y}
\eeq
differ only by some term proportional to identity.
\end{lemma}
\emph{Proof}. Obvious by differentiating the unitary condition \eqref{unit:general} and using regularity. \finproof

\bigskip

Now, we would like to implement Idzumi's method in the case of bivariate $R$-matrices and try to build the full solution for the $R$ matrix starting from the minimum possible knowledge.
Consider the braided Yang--Baxter equation \eqref{YBE} and expand the $R$-matrix as
\begin{align} \label{eq:R_2_variables_taylor}
\check R(x,y) = \sum_{i=0}^\infty \sum_{j=0}^\infty \check R^{(ij)} x^i y^j \,.
\end{align}
\begin{thm}
Let $\check R$ be an analytical solution of the braided Yang--Baxter equation~\eqref{YBE}, with Taylor series expansion of the form~\eqref{eq:R_2_variables_taylor}. Suppose that $\check R(x,x) = \II \otimes \II$.
Then, one can reconstruct the full solution $\check R(x,y)$ once the matrix $\check R (x,0)$ is given.
\end{thm}

\emph{Proof}.
Using the identity equation $\check R (x,0) \check R (0,x) = \II \otimes \II$, one can compute $\check R (0,x)$, hence $\check R^{(i0)}$ and $\check R^{(0i)}$ are known for all $i \in \mathbb N$. Recall that $\check R^{(00)} = \II \otimes \II$. \\
Consider the braided Yang--Baxter equation~\eqref{YBE}, and take the coefficient of $x^m y^0 z^n$:
\begin{equation}
\sum_{ij} \check R_{12}^{(0,j)} \check R_{23}^{(m-i,n-j)} \check R_{12}^{(i,0)}
= 
\sum_{ij} \check R_{23}^{(i,0)} \check R_{12}^{(m-i,n-j)} \check R_{23}^{(0,j)} \,.
\end{equation}
This last equation can be rewritten as a recursion relation for $\check R^{(m,n)}$:
\begin{align}
\check R^{(m,n)}_{12} - \check R^{(m,n)}_{23} &= 
\sum_{i=1}^m \sum_{j=1}^n \left(\check R_{12}^{(0,j)} \check R_{23}^{(m-i,n-j)} \check R_{12}^{(i,0)} -
\check R_{23}^{(i,0)} \check R_{12}^{(m-i,n-j)} \check R_{23}^{(0,j)} \right) \nonumber \\
& \quad +
\sum_{i=1}^m \left(\check R_{23}^{(m-i,n)} \check R_{12}^{(i,0)} -
\check R_{23}^{(i,0)} \check R_{12}^{(m-i,n)} \right) \nonumber \\
& \quad +
\sum_{j=1}^n \left(\check R_{12}^{(0,j)} \check R_{23}^{(m,n-j)} -
\check R_{12}^{(m,n-j)} \check R_{23}^{(0,j)} \right) \,.
\end{align}
which is of the form
\begin{equation}
\check R^{(m,n)}_{12} - \check R^{(m,n)}_{23} = Q^{(m,n)} \,,
\label{eq:recursion_double}
\end{equation} 
where $Q^{(m,n)}$ depends only on terms of smaller order in $m,n$.
Notice that we need $m$ and $n$ to be non-zero, otherwise the equation is trivial. \\
The rest of the proof is similar to the one variable case.
\finproof

\begin{coro}
If $\check R (x,0)$ preserves the spin, that is
$[ \check R (x,0) , \fs^z \otimes \II + \II \otimes \fs^z ] =0$, 
then it is also true for the full solution: 
$[ \check R (x,y) , \fs^z \otimes \II + \II \otimes \fs^z ] =0$.
\end{coro}
We need to check that the recursion relation preserves this property. However, relation \eqref{eq:recursion_double} is exactly of the same type that the one that appears in the one variable case, and the proof follows.

\bigskip

To be complete, notice that the fact that $\check R (x,x) = \II \otimes \II$ also leads to the equation:
\[
\sum_{i=0}^n \check R^{(i,n-i)} = \delta_{n,0}\, \II \otimes \II \,.
\]

\subsection{Resolution of YBE by brute force for multiplicative $R$-matrices}

Although the resolution of the Yang--Baxter equation seems to be an impossible task in the general case, the property of $U(1)$-invariance implies strong constraints on the resulting equations, and allows one in some cases to compute directly the $R$-matrix when it is univariate. Let us start with a regular $R$-matrix of the form
\beq
R(u)=
\begin{pmatrix} 
f_{11}(u) & 0 & 0 & 0 & 0 & 0 & 0 & 0 & 0 \\
0 & f_{22}(u) & 0 & f_{24}(u) & 0 & 0 & 0 & 0 & 0 \\
0 & 0 & f_{33}(u) & 0 & f_{35}(u) & 0 & f_{37}(u) & 0 & 0 \\
0 & f_{42}(u) & 0 & f_{44}(u) & 0 & 0 & 0 & 0 & 0 \\
0 & 0 & f_{53}(u) & 0 & f_{55}(u) & 0 & f_{57}(u) & 0 & 0 \\
0 & 0 & 0 & 0 & 0 & f_{66}(u) & 0 & f_{68}(u) & 0 \\
0 & 0 & f_{73}(u) & 0 & f_{75}(u) & 0 & f_{77}(u) & 0 & 0 \\
0 & 0 & 0 & 0 & 0 & f_{86}(u) & 0 & f_{88}(u) & 0 \\
0 & 0 & 0 & 0 & 0 & 0 & 0 & 0 & f_{99}(u) 
\end{pmatrix}
\label{eq:Rmat19}
\eeq
where the functions $f_{ij}(u)$ are to be determined. \\
We impose the Yang--Baxter equation and $\check{R}'(1)=H$ for a given Hamiltonian $H$, and we set 
\beq
p_{ij} = f_{ij}'(1).
\eeq
 This leads to a set of equations that should be satisfied by the functions $f_{ij}$. Among these equations, we start from the relation
\begin{equation}
f_{44}(u) f_{42}(uv) f_{22}(v) = f_{11}(u) f_{42}(uv) f_{11}(v) - f_{42}(u) f_{11}(uv) f_{42}(v) \,.
\end{equation}
Since the right hand side is symmetric in the exchange $u \leftrightarrow v$, one gets the consistency condition 
$p_{22} \, f_{44}(u) = p_{44} \, f_{22}(u)$.

In the same way, one obtains the consistency condition $p_{66} \, f_{88}(u) = p_{88} \, f_{66}(u)$.

Moreover, one has also a relation of the type $F(u)\,F(v)=F(uv)$ for $F(u) = f_{24}(u)/f_{42}(u)$ and for $F(u) = f_{68}(u)/f_{86}(u)$, therefore 
\begin{equation}
\frac{f_{24}(u)}{u^{p_{24}}} = \frac{f_{42}(u)}{u^{p_{42}}}
\qquad \text{and} \qquad
\frac{f_{68}(u)}{u^{p_{68}}} = \frac{f_{86}(u)}{u^{p_{86}}} \,.
\end{equation}
The remaining equations imply the two following relations: 
\beq \frac{f_{35}(u)}{f_{53}(u)}\,v^{p_{42}-p_{24}}=\frac{f_{35}(uv)}{f_{53}(uv)}
\mb{and} \frac{f_{75}(u)}{f_{57}(u)}\,v^{p_{24}-p_{42}}=\frac{f_{75}(uv)}{f_{57}(uv)}, 
\eeq
from which one obtains $f_{53}(u) = \lambda_{53} \, f_{35}(u) \, u^{p_{42}-p_{24}}$ and $f_{57}(u) = \lambda_{57} \, f_{75}(u) \, u^{p_{24}-p_{42}}$, where $\lambda_{53}$ and $\lambda_{57}$ are (at that point) arbitrary parameters. 

Note that $f_{35}(u)$ and $f_{75}(u)$ cannot be identically zero since the parameters of the Hamiltonian are restricted to $t_1 \ne 0$, $t_2 \ne 0$.
It is then convenient to introduce equations that depend only on one variable. To this end, we derive the YBE with respect to the $v$ variable and we set $v=1$. Similarly, we derive the YBE with respect to the $u$ variable and we set $u=1$. In this way, we obtain a set of differential equations satisfied by the functions $f_{ij}(u)$. We note however that among these equations, a particular subset ${\cal P}$ is constituted by polynomial functional equations. 

Two types of models emerge: 
\begin{itemize}
\item[$\diamond$]
In the first case, the parameters of the Hamiltonian satisfy $s_1=s_2=0$, from which it follows $\lambda_{53} = \lambda_{57} = 0$. Plugging this constraints in the YBE leads immediately to the following relations:
\begin{equation}
\frac{f_{37}(u)}{u^{p_{37}}} = \frac{f_{73}(u)}{u^{p_{73}}} \,,\quad p_{77} \, f_{44}(u) = p_{77} \, f_{44}(u) \,,\quad p_{88} \, f_{77}(u) = p_{77} \, f_{88}(u)
\end{equation}
At this stage, it is necessary to implement the different models by specifying for each case the entries of the Hamiltonian (i.e. the $p_{ij}$), at least for the off-diagonal part, and solve case by case the remaining equations.
\item[$\diamond$]
In the second case, $s_1$ and/or $s_2$ are nonzero and one should have $p_{24}-p_{42} = p_{68}-p_{86}$.
We focus now on the polynomial equations, once $f_{33}$ and $f_{77}$ have been expressed in terms of the other functions. In the case where $p_{22}p_{88}=p_{66}p_{44}$ (which is satisfied by the ZF, IK and SpR models), we can determine quite easily simple relations for the functions $f_{11}$, $f_{44}$, $f_{88}$, $f_{99}$, $f_{35}$, $f_{75}$, and in particular $f_{42} = f_{86}$. One deduces that $p_{22}^2p_{77} = p_{44}^2p_{33}$ (case of ZF or IK models for which one can proceed further) or $p_{44}^2 f_{11}(u) = p_{88}^2 f_{99}(u)$ (SpR model).
\end{itemize}
This method is exhaustive although rather heavy. However it remains tractable for univariate $R$-matrices in the general case. 
In the bivariate case, it becomes too intricate and one needs to restrict with some symmetry assumptions.

\subsection{Spectral curves \textsl{\`a la Martins}}

In \cite{M13}, M.J. Martins revisited the problem of $U(1)$-invariant three-state vertex models, when the Boltzmann weights $W_{\alpha\beta}^{\gamma\delta}$ configurations break the parity-time reversal symmetry, i.e. $W_{\alpha\beta}^{\gamma\delta} \ne W_{\delta\gamma}^{\beta\alpha}$. More precisely, the Lax operator being given by 
\begin{equation}
L_i(z) = \sum_{\alpha,\beta,\gamma,\delta} W_{\alpha\beta}^{\gamma\delta} \, e_{\alpha\gamma}^{(q)} \otimes e_{\beta\delta}^{(i)}
\end{equation}
where the index $q$ denotes the quantum space and $i$ the auxiliary spaces ($i=1,...,N$), one investigates the solutions of a general two parametric Yang--Baxter equation 
\begin{equation}
R(x,y)\, L_1(x)\,  L_2(y) = L_2(y)\, L_1(x)\, R(x,y) \,.
\label{eq:YB_RLL}
\end{equation}
Under some symmetry requirements for the $L$-matrix, it is possible in a first step to present a general formula for the $R$-matrix entries in terms of those of the $L$-matrix, and in a second step to express the $L$-matrix in terms of only two functions $a(z)$ and $b(z)$. 

At this point, two  solutions naturally emerge, denoted \emph{main branch (MB)} and \emph{special branch (SB)}. For each of these branches, the two functions $a(z)$ and $b(z)$ satisfy some polynomial relation that defines an elliptic curve, whose degree depends on the considered branch
\footnote{There are some misprints in \cite{M13} in the elliptic curve equations.}:
\bea
0 &=& (\alpha\beta-1)\big(a^4+a^2b^2+b^4\big)^2
+ (2-\alpha\beta+\alpha^2+\beta^2)\big(\beta(a^4+a^2b^2+b^4)+ab\big)ab
\nonu
&&-(\alpha\beta-2)a^4+\beta^2a^2b^2-(2-\alpha\beta+\beta^2)b^4-2\beta ab-1,
\qquad (MB)\label{MB},\\
0 &=& (a^2+jb^2)\big(a^4+a^2b^2+b^4+\Lambda_4ab\big)
+ b^2-a^2,\qquad (SB)\label{SB},
\eea
where $a\equiv a(z)$, $b\equiv b(z)$, $j^2-j+1=0$, and $\alpha$, $\beta$, $\Lambda_4$ are free parameters.

We will show below that the branches share the same Hamiltonian, if one generalizes the point where the Hamiltonian is defined (see below). In that sense, it is enough to consider the special branch, and study what is called the generalized Bariev model in \cite{r1d3}. Let us also stress that the branches are the most natural, but not the full set of choices one can do in solving the equations: a detailed study of these equations could lead to new (marginal) integrable models.

Relaxing the symmetry constraints used in \cite{M13}, one can try to use this method to compute solutions of the Yang--Baxter equation~\eqref{eq:YB_RLL}. This is a system of cubic equations containing in total $129$ equations (the others being zero by conservation of spin), each equation being linear in the entries of $R$ and quadratic in the entries of $L$. By choosing a suitable set of equations, one can obtain the entries of $R$ in terms of the entries of $L$, up to a multiplicative constant set by normalizing some given entry of $R$ to $1$. 

The remaining equations can be seen as complementary equations for the entries of $L$. Plugging the computed entries of $R$ in these equations, such complementary equations can become very involved, and so, it is crucial to chose wisely the set of equations to use for determining $R$.

These final equations depend on all sorts of entries of $L(x)$ and $L(y)$. Untangling them may happen naturally, otherwise one can try to write the equation $f(L(x),L(y))=0$ as:
\begin{equation}
  f(L(x),L(y)) = P(L(x)) Q(L(y)) - P(L(y)) Q(L(x)) = 0 \,.
\end{equation}
When this is possible, one can then write $P(L(x)) = c\, Q(L(x))$, hence the emergence of elliptic curves depending on some constants in the computation.
Notice that we should make sure that $Q(L(x))$ does not vanish.

\section{Hamiltonians and $R$-matrices}
\label{sect:resu}

We consider here only Hamiltonians that lead to non-trivial scattering matrices ($S\neq \pm 1$). Indeed, when the scattering matrix is trivial, the Bethe equations become also trivial, and the CBA obviously fails to provide a complete spectrum of the Hamiltonian. 
Note however that there are some cases where the scattering is trivial, but one can nevertheless construct a $R$-matrix. The status of the corresponding Hamiltonians concerning integrability remains unclear, since CBA does not provide the complete spectrum, but one still gets a transfer matrix $t(u)$ that commutes for different values of the spectral parameter.

The Hamiltonians of \cite{r1d3} that have non-trivial scattering matrix are the four 19-vertex, two 17-vertex and one 14-vertex. Before 
presenting them, we simplify them using a twist procedure.

\subsection{19-vertex Hamiltonians}

\subsubsection{Zamolodchikov--Fateev Hamiltonian} 

We found in \cite{r1d3} an expression for the Zamolodchikov--Fateev Hamiltonian $H_{\textrm{ZF}}$ that contained a supplementary parameter $\tau_p$ with respect to the original one \cite{ZF}. For $\tau_p=-1$, this Hamiltonian is related to the one based on $\cU_q(B_1^{(1)})$ given in \cite{J86} by 
\begin{equation}
H_{\cU_q(B_1^{(1)})}(1/k^2) = H_{\textrm{ZF}}(k)\vert_{\tau_p=-1} + (\II\otimes e_{22}-e_{22}\otimes\II) + 2(\II\otimes e_{33}-e_{33}\otimes\II) \,,
\label{eq:HZF}
\end{equation}
where the $R$-matrix of $\cU_q(B_1^{(1)})$ is normalized by $R_{11}^{11}=1$. \\
In fact, one can remove this parameter by considering the transformation 
\begin{equation}
\wt H_{\textrm{ZF}} = F_{12}(\tau_p) H_{\textrm{ZF}} F_{12}^{-1}(\tau_p)
\label{eq:conjZF}
\end{equation}
where $F_{12}(\tau_p) = g \, \exp({\alpha\, \fs^z}) \otimes g \, \exp({-\alpha\, \fs^z})$ with $g=\diag\,(1,\tau_p^{-1/2},1)$ and $e^\alpha=\tau_p^{1/2}$. The explicit expression of $\wt H_{\textrm{ZF}}$ is formally obtained by setting $\tau_p=1$ in $H_{\textrm{ZF}}$. \\
It follows that $\wt H_{\textrm{ZF}}(k)$ is related to $H_{\cU_q(B_1^{(1)})}(1/k^2)$, up to the telescopic terms of \eqref{eq:HZF}, by the transformation of the type \eqref{eq:conjZF} with $F_{12}(-1)=\diag(1,-1,-1,1,-1,-1,-1,1,1)$. 
Therefore the $R$-matrix from which one can deduce $\wt H_{\textrm{ZF}}$ is obtained by twisting the $R$-matrix of $\cU_q(B_1^{(1)})$ by this $F_{12}(-1)$. It can be checked that the obtained $R$-matrix satisfies the Yang-Baxter equation.
The explicit expression of the $R$-matrix is as follows:
\begin{equation}
R(u) = \left( \begin {array}{ccccccccc} 
1 & 0 & 0 & 0 & 0 & 0 & 0 & 0 & 0 \\
0 & b(u) & 0 & c_-(u) & 0 & 0 & 0 & 0 & 0 \\
0 & 0 & f(u) & 0 & d_-(u) & 0 & h_-(u) & 0 & 0 \\
0 & c_+(u) & 0 & b(u) & 0 & 0 & 0 & 0 & 0 \\
0 & 0 & d_+(u) & 0 & g(u) & 0 & d_-(u) & 0 & 0 \\
0 & 0 & 0 & 0 & 0 & b(u) & 0 & c_-(u) & 0 \\
0 & 0 & h_+(u) & 0 & d_+(u) & 0 & f(u) & 0 & 0 \\
0 & 0 & 0 & 0 & 0 & c_+(u) & 0 & b(u) & 0 \\
0 & 0 & 0 & 0 & 0 & 0 & 0 & 0 & 1 
\end{array}\right)
\label{eq:RmatZF}
\end{equation}
the entries of the $R$-matrix being given by
\begin{align}
& b(u)=-\frac{(u^2-1)k^2}{k^4u^2-1} \,,\qquad c_\pm(u)=u^{\pm 1}\,\frac{u(k^4-1)}{k^4u^2-1} \,,\qquad f(u)=\frac{(u^2-k^2)(u^2-1)k^2}{(k^4u^2-1)(k^2u^2-1)} \\
& d_\pm(u)=-u^{\pm 1}\,\frac{uk(k^4-1)(u^2-1)}{(k^4u^2-1)(k^2u^2-1)} \,,\qquad h_\pm(u)=u^{\pm 2}\,\frac{u^2(k^4-1)(k^2-1)}{(k^4u^2-1)(k^2u^2-1)} \\
& g(u)=\frac{k^4u^4+u^2(k^2+1)(k^2+k-1)(k^2-k-1)+k^2}{(k^4u^2-1)(k^2u^2-1)} 
\end{align}

\subsubsection{Izergin--Korepin Hamiltonian} 

We found in \cite{r1d3} an expression for the Izergin--Korepin Hamiltonian $H_{\textrm{IK}}$ that contained a supplementary parameter $\tau'_p$, see formula (5.7) therein. This parameter can be removed by considering the transformation 
\begin{equation}
\wt H_{\textrm{IK}} = F_{12}(\tau'_p) H_{\textrm{IK}} F_{12}^{-1}(\tau'_p)
\label{eq:conjIK}
\end{equation}
where $F_{12}(\tau'_p) = \exp({\alpha\, \fs^z}) \otimes \exp({-\alpha\, \fs^z})$ with $e^\alpha={\tau'_p}^{1/2}$. The explicit expression of $\wt H_{\textrm{IK}}$ is formally obtained by setting $\tau'_p=1$ in $H_{\textrm{IK}}$. One gets the Hamiltonian based on the $R$-matrix of $\cU_q(A_2^{(2)})$ given in \cite{J86} and normalized such that $R_{11}^{11}=1$. One recovers also the Izergin--Korepin $R$-matrix of the Shabat--Mikhailov model \cite{IK}, after exchanging the roles of the states $|0\rangle$ and $|1\rangle$, and taking into account into the $R$-matrix the existence of telescopic terms and a gauge transformation between the corresponding Hamiltonians, see section \ref{subsect:transfo}.

The $R$-matrix has the same shape as in \eqref{eq:RmatZF} but its entries are now given by
\begin{align}
& b(u)=\frac{k(u-1)}{u-k^2} \,,\quad d_-(u)=k^2\,d(u) \,, \quad d_+(u)=-u\,d(u) \,, \quad d(u)=\frac{k^{1/2}(1-k^2)(u-1)}{(u+k^3)(u-k^2)} \\
& c_-(u)=\frac{1-k^2}{u-k^2} \,,\qquad c_+(u)=uc_-(u) \,,\qquad f(u)=\frac{k^2(u+k)(u-1)}{(u+k^3)(u-k^2)} \\
& g(u)=\frac{k(u+k^3)(u-1)+u(k^3+1)(1-k^2)}{(u+k^3)(u-k^2)} \\
& h_-(u)=\frac{(u+k^3+k^2(u-1))(1-k^2)}{(u+k^3)(u-k^2)} \,,\qquad h(u)_+=\frac{(u+k^3-k(u-1))(1-k^2)}{(u+k^3)(u-k^2)} 
\end{align}

\subsubsection{Generalized Bariev Hamiltonian} \label{sect:model25} 

As in the previous cases, one can simplify further the Hamiltonian found in \cite{r1d3}, see formula (5.13) therein. First of all, we 
perform the following change of variables: 
\begin{equation}
\phi = \frac{\xi}{J_0\tau_p\mu^{1/2}} \qquad \text{and} \qquad \psi = \frac{\xi\theta}{J_0^3\tau_p\mu^{3/2}} 
\end{equation}
and define $\upsilon$ as
\begin{equation}
-4\xi\upsilon = \phi^2 - \phi\psi + \psi^2 \,,
\label{eq:def_delta}
\end{equation}
where $J=-1/J_0^2$ and $\xi$ is some normalization constant introduced for later convenience (although it may be set to one for the moment, it is useful for taking some limits, see below). Then, using the transformation \eqref{eq:gradation} with $\alpha=(-J^2\mu)^{-1/4}$, one gets an Hamiltonian that depends only on $\phi$, $\psi$ and $\xi$:
\begin{equation}
H_{\textrm{GB}} = 
\begin{pmatrix}
-\upsilon & 0 & 0 & 0 & 0 & 0 & 0 & 0 & 0 \\
0 & 0 & 0 & \phi & 0 & 0 & 0 & 0 & 0 \\
0 & 0 & -\upsilon - J^2\xi & 0 & \phi & 0 & \xi & 0 & 0 \\
0 & \psi & 0 & 0 & 0 & 0 & 0 & 0 & 0 \\
0 & 0 & -J^2 (\psi-\frac{\xi^2}{\phi}) & 0 & \upsilon - \xi & 0 & \psi-\frac{\xi^2}{\phi} & 0 & 0 \\
0 & 0 & 0 & 0 & 0 & 0 & 0 & \psi & 0 \\
0 & 0 & \xi & 0 & -J \phi & 0 & -\upsilon - J \xi & 0 & 0 \\
0 & 0 & 0 & 0 & 0 & \phi & 0 & 0 & 0 \\
0 & 0 & 0 & 0 & 0 & 0 & 0 & 0 & -\upsilon 
\end{pmatrix} \,.
\label{htilde24}
\end{equation}
As stated in \cite{r1d3}, this Hamiltonian is a generalization of the one found by Alcaraz and Bariev \cite{AB}, further corrected in \cite{AN}.

\paragraph{Relation with the Hamiltonian of the main branch at the special point $H_{\textrm{MB}_0}$.}
The Hamiltonian $H_{\textrm{GB}}$ can be related to the one $H_{\textrm{MB}_0}$ obtained in \cite{M13} in the case of the main branch. More precisely, the two-site Hamiltonian $H_{\textrm{MB}_0}$ is defined as 
\begin{equation}
H_{\textrm{MB}_0} = P \; \frac{\partial}{\partial x} L(x)\Big\vert_{x=x_0}
\end{equation}
where the point $x_0$ is chosen such that $a(x_0)=1$, $b(x_0)=0$ (constraints satisfied by the corresponding elliptic curve). In order to compute explicitly $H_{\textrm{MB}_0}$, one differentiates the elliptic curve $\gamma(a(x),b(x))=0$ and obtains $b^\prime(x_0)$ as a function of $a^\prime(x_0)$:
\begin{equation}
b^\prime(x_0) = \frac{-4 \alpha}{\alpha^2-\alpha \beta + \beta^2} \, a^\prime(x_0) \,,
\end{equation}
where $\alpha$ and $\beta$ are the two constants entering in the definition of the curve, see \eqref{MB}. \\
The Hamiltonian $H_{\textrm{MB}_0}$ then takes the form \cite{M13}:
\begin{equation}
\label{eq:H_MB_0}
H_{\textrm{MB}_0} = 
\begin{pmatrix}
1 & 0 & 0 & 0 & 0 & 0 & 0 & 0 & 0 \\
0 & 0 & 0 & -\beta \rho & 0 & 0 & 0 & 0 & 0 \\
0 & 0 & 1+J_0^2 \rho & 0 & -J_0^2 \eta \rho & 0 & \rho & 0 & 0 \\
0 & - \alpha \rho & 0 & 0 & 0 & 0 & 0 & 0 & 0 \\
0 & 0 & - J_0^2 \eta \rho & 0 & -1 - \rho & 0 & -\eta \rho & 0 & 0 \\
0 & 0 & 0 & 0 & 0 & 0 & 0 & - \alpha \rho & 0 \\
0 & 0 &  \rho & 0 & -\eta \rho & 0 & 1+J_0^{-2}\rho & 0 & 0 \\
0 & 0 & 0 & 0 & 0 & -\beta \rho & 0 & 0 & 0 \\
0 & 0 & 0 & 0 & 0 & 0 & 0 & 0 & 1
\end{pmatrix}\ ,
\end{equation}
where
\begin{equation}
\rho = \frac{4}{\alpha^2 - \alpha \beta + \beta^2} \qquad \text{and} \qquad \eta = -J_0^{-1} \sqrt{\alpha\beta-1} \,.
\end{equation}
Then the Hamiltonian $H_{\textrm{GB}}$ can be related to $H_{\textrm{MB}_0}$ by the gauge transformation
\beq
H_{\textrm{GB}} = (F \otimes F)\, H_{\textrm{MB}_0} \,(F \otimes F)^{-1}
\eeq
where 
\beq
F = \diag\left(1,\sqrt{\frac{J_0^2\eta}{\beta}},1\right) 
\label{eq:twistbariev}
\eeq
and the correspondence between the parameters is  
\beq
\upsilon = -1 \;, \qquad \alpha = \frac{\psi}{\xi} \;, \qquad \beta = \frac{\phi}{\xi} \;, \qquad \rho = \xi
\eeq
Therefore the $R$-matrix from which the Hamiltonian $H_{\textrm{GB}}$ can be derived is obtained by twisting by $F$, formula \eqref{eq:twistbariev}, the $R$-matrix corresponding to the main branch of ref. \cite{M13}.

\paragraph{Relation with the Hamiltonian of the special branch at a generic point $H_{\textrm{SB}}$.}
As mentioned in section \ref{sect:setup}, when the $R$ matrix satisfies $\check R(x,x)=\II$, one can also define the Hamiltonian for bivariate $R$ matrices by $H = \left. \partial_x \check R(x,y) \right|_{x=y}$ keeping the spectral parameter $x$ free\footnote{One can check by direct calculation that the $R$-matrices in \cite{M13} are regular, both for MB and SB.}. In this way, one introduce a new parameter $a(x)$ (or equivalently $b(x)$) that was set to 1 in the construction of \cite{M13} (since $a(x_0)=1$).
We now present the corresponding Hamiltonians, both for the main branch and for the special branch.

In the case of the special branch, defining $H_{\textrm{SB}} = \left. \partial_x \check R(x,y) \right|_{x=y}$ where $\check R$ is given by the formulae (119-121) of \cite{M13}, one gets
\begin{align}
H_{\textrm{SB}} &= 
h_a (E_{00} \otimes E_{00}+E_{22} \otimes E_{22}) + h_{\bar h} E_{00} \otimes E_{22} + h_g E_{11} \otimes E_{11} + h_h E_{22} \otimes E_{00} \nonumber \\
& + h_f (E_{02} \otimes E_{20}+E_{20} \otimes E_{02}) + h_b (E_{10} \otimes E_{01}+E_{12} \otimes E_{21}) + h_{\bar b} (E_{01} \otimes E_{10}+E_{21} \otimes E_{12}) \nonumber \\
& + h_d (E_{12} \otimes E_{10}+E_{21} \otimes E_{01}) + h_{\bar d} (E_{01} \otimes E_{21}+E_{10} \otimes E_{12}) \,,
\label{eq:HSB}
\end{align}
and we set $h_f=\xi$, $h_b=\psi$, $h_{\bar b}=\phi$, $h_a=-\upsilon$ (expressed in terms of the two functions $a(x)$ and $b(x)$). From the expressions of the entries of the $R$-matrix, one obtains immediately $h_h = -\upsilon - J\xi$ and $h_{\bar h} = -\upsilon - J^2\xi$. Using then a symbolic computation program, it can be checked that $H_g = \upsilon + \xi$ up to the curve $\gamma$. 
Consider then $F$, a diagonal matrix of the kind $g \otimes g$ where $g=\diag(1,\sqrt{\zeta},1)$, and compute $F H_{\textrm{SB}} F^{-1}$. The entries that are changed are those corresponding to $h_d$ and $h_{\bar d}$ with a factor $\zeta$ for the $_{02}^{11}$ and $_{20}^{11}$ entries and a factor $\zeta^{-1}$ for the $_{11}^{02}$ and $_{11}^{20}$ entries. In order to get further in the identification with the generalized Bariev Hamiltonian, one has to impose
\begin{equation}
\zeta^{-1} h_{\bar d} = \phi \,, \qquad
\zeta^{-1} h_d = - J \phi \,, \qquad
\zeta h_{\bar d} = -J^2 (\psi - \frac{\xi^2}{\phi}) \,, \qquad
\zeta h_d = \psi - \frac{\xi^2}{\phi} \,.
\end{equation}
Hence we set $\displaystyle \zeta = \frac{h_{\bar d}}{\phi}$ and the last equation becomes
\[
h_d h_{\bar d} = \psi \phi - \xi^2 \qquad \textrm{up to the curve}\  \gamma\ ,
\]
which can be checked by a symbolic computation program.
\\
Finally, equation~\eqref{eq:def_delta} can also be checked, up to the curve $\gamma$.
The factor $\xi$ is just a normalization and therefore there are only two genuine parameters $\psi/\epsilon$ and $\phi/\epsilon$, which will depend on the two parameter of the model $\zeta$ and $\Lambda_4$ (one can prove that they are nonzero and independent by computing the Wronskian).

\paragraph{Relation with the Hamiltonian of the main branch at a generic point $H_{\textrm{MB}}$.}
In the case of the main branch, defining $H_{\textrm{MB}} = \left. \partial_x \check R(x,y) \right|_{x=y}$ where $\check R$ is given by the formulae (115-118) of \cite{M13}, one gets for the two-site Hamiltonian a matrix similar to \eqref{eq:HSB}, where $H_c = 0$ and all other entries are very complicated (the smallest one $H_a$ occupying five lines in Mathematica output and the biggest one $H_g$ occupying more than six pages). The reasoning follows the same lines as in the special branch case, and finally the Hamiltonian $H_{\textrm{GB}}$ also appears as a twist of the Hamiltonian $H_{\textrm{MB}}$. 

It follows that the same Hamiltonian $H_{\textrm{GB}}$ can be derived in three different ways by considering either the $R$-matrix of the main branch (at the special point $x_0$ or at a generic point) or the $R$-matrix of the special branch at a generic point.
Hence the main branch and the special branch of \cite{M13} share the same conserved quantities.

Therefore the $R$-matrix of the generalized Bariev model is the one given in \cite{M13}, where one can restrict oneself to the special branch case. Hence, the $R$-matrix takes the form
\begin{align}
R(x,y) &= 
r_a (E_{00} \otimes E_{00}+E_{22} \otimes E_{22}) + r_b (E_{00} \otimes E_{11}+E_{22} \otimes E_{11}) + r_{\bar b} (E_{11} \otimes E_{00}+E_{11} \otimes E_{22}) \nonumber \\
& + r_f (E_{00} \otimes E_{22}+E_{22} \otimes E_{00}) + r_g E_{11} \otimes E_{11} \nonumber \\
& + r_c (E_{01} \otimes E_{10}+E_{10} \otimes E_{01}+E_{12} \otimes E_{21}+E_{21} \otimes E_{12}) + r_h E_{02} \otimes E_{20} + r_{\bar h} E_{20} \otimes E_{02} \nonumber \\
& + r_d (E_{01} \otimes E_{21}+E_{12} \otimes E_{10}) + r_{\bar d} (E_{10} \otimes E_{12}+E_{21} \otimes E_{01}) \,,
\label{eq:Rmartins}
\end{align}
where the entries are given in terms of two functions $a(z)$ and $b(z)$ that satisfy equation \eqref{SB}:
\begin{align}
r_a(x,y) &= \frac{a(x)a(y)}{a(y)^2+jb(y)^2} + \frac{jb(x)b(y)}{a(x)^2+jb(x)^2} \\
r_b(x,y) &= \frac{b(x)a(y)}{a(y)^2+jb(y)^2} - \frac{a(x)b(y)}{a(x)^2+jb(x)^2} \\
r_{\bar b}(x,y) &= \frac{jb(x)a(y)}{a(x)^2+jb(x)^2} - \frac{ja(x)b(y)}{a(y)^2+jb(y)^2} \\
r_d(x,y) &= 
j\frac{b(x)a(y)(a(y)^2+jb(y)^2)-a(x)b(y)(a(x)^2+jb(x)^2)}{ja(x)a(y)+b(x)b(y)(a(x)^2+jb(x)^2)(a(y)^2+jb(y)^2)}
\\
r_f(x,y) &= r_d(x,y)\;\frac{b(x)a(y)(a(x)^2+jb(x)^2)(a(y)^2+jb(y)^2)-j^2a(x)b(y)}{(a(x)^2+jb(x)^2)(a(y)^2+jb(y)^2)} \\
r_g(x,y) &= -r_d(x,y)\;\frac{r_f(y,x)+r_a(y,x)}{r_d(y,x)} \;, \qquad r_{\bar d}(x,y) = jr_d(x,y) \\
r_h(x,y) &= r_a(x,y) + j^{-1}r_f(x,y) \;, \qquad r_{\bar h}(x,y) = r_a(x,y) + jr_f(x,y)
\end{align}
the $R$-matrix being normalized such that $r_c(x,y)=1$.

\subsubsection{Generalized SpR Hamiltonian\label{sect:SpR}} 

Performing the change of variable $\displaystyle\theta\tau_p^2=\theta_0=\frac{qt_p^2}{p^3}$, the transformation \eqref{eq:gradation} acting on $H_{red}$ (see eq. (5.19) of \cite{r1d3}) with $\alpha={\tau_p}^{-1/2}$ leads to an Hamiltonian which does not depend any longer on $\tau_p$, with $\delta_0=\tau_3^2-\tau_3+1$. The explicit expression of $\wt H_{\textrm{SpR}}$ is formally obtained by setting $\tau_p=1$, $\theta\to\theta_0$, $\delta\to\delta_0+\theta_0$ in $H_{red}$. \\

The method of finding the $R$-matrix by brute force when considering multiplicative spectral parameter implementation can be used in the case of the SpR model. Referring to the ``algorithm'' explained above, one is led to the following constraint: $p_{44}^2 f_{11}(u) = p_{88}^2 f_{99}(u)$, that is here $f_{11}(u) = \tau_3^2 f_{99}(u)$. Plugging this last equation in the polynomial set ${\cal P}$ leads to impose specific values of the parameters of the Hamiltonian, namely $\tau_3^2=1$ and $\theta_0=\tau_3^2-\tau_3+1$. One concludes that the SpR model does not admit a $R$-matrix with only one multiplicative spectral parameter for generic values of the parameters $\theta_0$ and $\tau_3$. 

It is of course tempting to test the case of bivariate $R$-matrices. Unfortunately, the SpR Hamiltonian shows very few symmetry and it is too intricate to deal with the general case. Although the resolution of this case remains open, it implies that if a (bivariate) $R$-matrix exists for this Hamiltonian, it looks certainly very intricate.

Moreover, we checked that the spectrum (and indeed the multiplicities) of the Hamiltonian was obtained from CBA in the case of a chain of length two.

It leaves the question open whether "it always exists an $R$-matrix when an Hamiltonian is solvable by CBA"? or in other words "does CBA implies ABA?"

\subsection{17-vertex Hamiltonians}

\subsubsection{Generalized ``Special Branch'' Hamiltonian} 

Performing the transformation \eqref{eq:gradation} acting on $\wt H$ (eq. (5.25) of \cite{r1d3} suitably normalized) with $\alpha=(-JQ)^{-1/4}$ leads to an Hamiltonian which does not depend any longer on $Q$: 
\begin{align}
H_{17} = &-\Lambda (E_{00} \otimes E_{00} + E_{00} \otimes E_{22} - E_{11} \otimes E_{11} + E_{22} \otimes E_{00} + E_{22} \otimes E_{22}) \nonumber \\
&- J (E_{01} \otimes E_{10} + E_{01} \otimes E_{21} + E_{10} \otimes E_{12} + E_{21} \otimes E_{21}) \nonumber \\
&+ E_{10} \otimes E_{01} + E_{12} \otimes E_{10} + E_{12} \otimes E_{21} + E_{21} \otimes E_{01}
\end{align}
which is directly related to the Hamiltonian $H_{\textrm{SB0}}$ of \cite{M13}, where $\displaystyle\Lambda = \frac{\Upsilon}{4J\sqrt{-Q}}$\,. \\
Note that the Hamiltonian $H_{17}$ can be obtained as a limit of the Hamiltonian $H_{\textrm{GB}}$. Indeed, if we take the limit $\xi=0$ in the Hamiltonian $H_{\textrm{GB}}$ and set 
\begin{equation*}
\upsilon = \Lambda \;,\quad \phi = -J^2 \;,\quad \psi = 1 \;,
\end{equation*}
which satisfy equation~\eqref{eq:def_delta}, we obtain the Hamiltonian $H_{17}$ for the value $J \to J^2$.

\subsubsection{$17V_2$ Hamiltonian} 

Performing the change of variable $\displaystyle\theta\tau_p^2=\theta_0=\frac{qt_p^2}{p^3}$, the transformation \eqref{eq:gradation} acting on $H_{red}$ (eq. (5.33) of \cite{r1d3}) with $\alpha={\tau_p}^{-1/2}$ leads to an Hamiltonian $\wt H_{\textrm{17V$_2$}}$ which does not depend any longer on $\tau_p$. The explicit expression of $\wt H_{\textrm{17V$_2$}}$ is formally obtained by setting $\tau_p=1$ in $H_{red}$. 

It can be checked that this Hamiltonian satisfies the Hecke relations, and therefore one can derive the $R$-matrix using a Baxterization procedure. Explicitly one obtains:
\begin{equation} 
R(z) = \begin{pmatrix}
z-{\frac {\theta_0}{z}} & 0 & 0 & 0 & 0 & 0 & 0 & 0 & 0 \\ 
0 & (z-\frac{1}{z})\theta_0 & 0 & \frac {1-\theta_0}{z} & 0 & 0 & 0 & 0 & 0 \\ 
0 & 0 & (z-\frac{1}{z})\theta_0 & 0 & \frac{1}{z}-z & 0 & \frac {1-\theta_0}{z} & 0 & 0 \\ 
0 & z (1-\theta_0) & 0 & z-\frac{1}{z} & 0 & 0 & 0 & 0 & 0 \\ 
0 & 0 & 0 & 0 & z-{\frac {\theta_0}{z}} & 0 & 0 & 0 & 0 \\ 
0 & 0 & 0 & 0 & 0 & (z-\frac{1}{z})\theta_0 & 0 & \frac {1-\theta_0}{z} & 0 \\ 
0 & 0 & z (1-\theta_0) & 0 & z-\frac{1}{z} & 0 & z-\frac{1}{z} & 0 & 0 \\ 
0 & 0 & 0 & 0 & 0 & z (1-\theta_0) & 0 & z-\frac{1}{z} & 0\\ 
0 & 0 & 0 & 0 & 0 & 0 & 0 & 0 & z-{\frac {\theta_0}{z}}
\end {pmatrix} 
\label{R17V2}
\end{equation}
To the best of our knowledge, this $R$-matrix is new.

\subsection{$14V$ model} 

The transformation \eqref{eq:gradation} acting on $H_{red}$ (eq. (5.35) of \cite{r1d3}) with $\alpha={\tau_p}^{-1/2}$ leads to an Hamiltonian which does not depend any longer on $\tau_p$: 
\begin{equation}
H_{14V} = E_{01} \otimes E_{10} + E_{01} \otimes E_{21} - E_{21} \otimes E_{01} + E_{12} \otimes E_{21} + E_{02} \otimes E_{20} + \sum_{i,j} v_{ij} E_{ii} \otimes E_{jj}
\end{equation}
where the non vanishing $v_{ij}$ are $v_{02}=v_{20}=1/2$, $v_{11}=1$, $v_{12}=3/2$, $v_{21}=\xi-3/2$, $v_{22}=\xi$.

Although this Hamiltonian is very simple, one cannot construct a suitable $R$-matrix. More precisely, the following result can be proved by a brute force calculation: it is not possible to find a univariate or a bivariate $R$-matrix satisfying the YBE unless one chooses $\xi=2$. In that case, the obtained Hamiltonian $H_{14V}\vert_{\xi=2}$ is a particular case of the Hamiltonian $H_{17V_2}$ when $\theta_0=0$, hence the corresponding $R$-matrix is given by \eqref{R17V2} with $\theta_0=0$.

Thus, we have a model whose Hamiltonian is not built from an $R$-matrix, but on which one can perform (at least partially) the CBA. Note that  the pseudo-vacuum and the pseudo-plump are here both necessary (see appendix). However direct calculations for $L=2$ and $L=3$ show that a third reference state is needed for completeness. It \emph{could} be an example of a model solvable by CBA but not by ABA. However, a more detailed study on completeness of the CBA is needed before reaching such a conclusion.

\section*{Acknowledgements}
T.F. was supported by ANR Project
DIADEMS (Programme Blanc ANR SIMI1 2010-BLAN-0120-02).

\appendix

\section{CBA for the second reference state\label{CBA-e1}}

Given a $U(1)$-invariant Hamiltonian $H$, see eqs. \eqref{eqH1} and \eqref{eqH2}, we are interested here in deriving the CBA when considering the second reference state. Let us recall briefly the ansatz for the first reference state (pseudo-vacuum).

\subsection{CBA on the pseudo-vacuum}
 Since the $S^z$ component of the total spin commutes with the Hamiltonian, one can decompose the space of states into subspaces ${\cal V}_M$ with given $S^z$-eigenvalue $M$. The subspace ${\cal V}_0$ ($M=0$) is one-dimensional with basis given by the eigenvector $|\Omega\rangle = \bigotimes_{i=1}^L |0\rangle$, called the pseudo-vacuum, corresponding to the eigenvalue $Lv_{00}$. A basis of states in ${\cal V}_M$ with a given number $M$ of pseudo-excitations is obtained by acting with the raising operator on the pseudo-vacuum such that
\begin{equation}
\vert x_1,\dots,x_M \rangle = \underbrace{|0\rangle \otimes \dots \otimes |0\rangle}_{x_1-1} \otimes |m_1\rangle \otimes \underbrace{|0\rangle \otimes \dots \otimes |0\rangle}_{x_{m_1+1}-x_{m_1}-1} \otimes |m_2\rangle \otimes \underbrace{|0\rangle \otimes \dots \otimes |0\rangle}_{x_{m_1+m_2+1}-x_{m_1+m_2}-1} \otimes |m_3\rangle \otimes \dots 
\label{eq:elemstate}
\end{equation}
where $1\leq x_1\leq x_2\leq...\leq x_M\leq L$. The $x_j$'s are the locations of the pseudo-excitations along the chain, and $m_k \in \{1,2\}$ such that $\sum m_k = M$. For $j=1+m_1+\dots+m_{k-1}$, one has $m_k=2$ if $x_{j+1}=x_j$ and $m_k=1$ otherwise. \\

An eigenstate $\Psi_M$ for the Hamiltonian $H$ in ${\cal V}_M$ is given by suitable linear combinations of the elementary states \eqref{eq:elemstate} with coefficients $a(x_1,\dots,x_M)$, which are complex-valued functions to be determined:
\begin{equation}\label{eq:psiM}
\Psi_M = \sum_{1 \le x_1 \le \dots \le x_M \le L} a(x_1,\dots,x_M) \vert x_1,\dots,x_M \rangle,
\end{equation}
and we assume a plane wave decomposition for the functions $a(x_1,\dots,x_M)$:
\begin{equation}
a(x_1,\dots,x_M) = \sum_{\sigma \in {\mathfrak S}_M} A_\sigma^{(j_1,\dots,j_P)}(k_1,\dots,k_M) \exp\left(\sum_{n=1}^M ik_{\sigma(n)} x_n \right) = \sum_{\sigma \in {\mathfrak S}_M} A_\sigma^{(j_1,\dots,j_P)}(\vec{k}) e^{i\vec{k_\sigma}\cdot\vec{x}}.
\label{eq:planewave}
\end{equation}
Here ${\mathfrak S}_M$ is the permutation group of $M$ elements and $A_\sigma^{(j_1,\dots,j_P)}(k_1,\dots,k_M)$ are functions on the symmetric group algebra depending on the Bethe roots $k_n$ to be determined by the so-called Bethe ansatz equations. The indices $(j_1,\dots,j_P)$ correspond to double excitations, i.e. indices such that $x_{j_k+1}=x_{j_k}$ for $k=1,\dots,P$. 

The energy of the eigenstate $\Psi_M$ is then given by
\begin{equation}
E_M = Lv_{00} + M(v_{01}+v_{10}-2v_{00}) + \sum_{n=1}^M (q\,e^{ik_{n}}+p\,e^{-ik_{n}})  
\label{eq:energie0}
\end{equation}
and the Bethe equations that determine the Bethe roots $k_n$ are (see proposition 3.1 of \cite{r1d3} for an explicit expression of the $S$-matrix)
\begin{equation}
e^{ik_{j}L} = \prod_{n \ne j} S(k_n,k_j)\,, \qquad j=1,...,M.
\label{eq:BAE}
\end{equation}

\subsection{CBA on the pseudo-plump}
There exists another one-dimensional subspace, ${\cal V}_{2L}$, with eigenvector $|\widetilde\Omega\rangle = \bigotimes_{i=1}^L |2\rangle$ corresponding to the eigenvalue $Lv_{22}$. This eigenvector is another possible reference state, that we call the pseudo-plump by opposition to the pseudo-vacuum, on which one can develop the CBA method. The existence of the pseudo-plump is just the reflection of the charge conjugation transformation, $h_{i_1\;i_2}^{j_1\;j_2} \to h_{2-i_1\;2-i_2}^{2-j_1\;2-j_2}$, or in terms of the parameters of the Hamiltonian: 
\begin{equation}
v_{ij} \leftrightarrow v_{2-i,2-j} \,, \;\; p\leftrightarrow s_3 \,, \;\; q\leftrightarrow t_3 \,, \;\; t_1\leftrightarrow t_2 \,, \;\; s_1\leftrightarrow s_2 \,, \;\; t_p\leftrightarrow s_p.
\label{eq:transfoC}
\end{equation}
In most cases, there is no need to perform this second CBA, because all eigenstates can be obtained from the pseudo-vacuum. However, there are cases where both CBA are needed to get a complete set.

We choose now as a basis of states in ${\cal V}_N$ with a given number $N$ of pseudo-holes (note that $N=2L-M$), the states obtained by acting with the lowering operator on the pseudo-plump $|\widetilde\Omega\rangle$ such that
\begin{equation}
\vert y_1,\dots,y_N \rangle = \underbrace{|2\rangle \otimes \dots \otimes |2\rangle}_{y_1-1} \otimes |m_1\rangle \otimes \underbrace{|2\rangle \otimes \dots \otimes |2\rangle}_{y_{m_1+1}-y_{m_1}-1} \otimes |m_2\rangle \otimes \underbrace{|2\rangle \otimes \dots \otimes |2\rangle}_{y_{m_1+m_2+1}-y_{m_1+m_2}-1} \otimes |m_3\rangle \otimes \dots 
\label{eq:elemstate2}
\end{equation}
where $1\leq y_1\leq y_2\leq...\leq y_N\leq L$. The $y_j$'s are the locations of the pseudo-holes along the chain, and $m_k \in \{1,2\}$ such that $\sum m_k = N$. For $j=1+m_1+\dots+m_{k-1}$, one has $m_k=2$ if $y_{j+1}=y_j$ and $m_k=1$ otherwise. \\

An eigenstate $\Psi_N$ for the Hamiltonian $H$ in ${\cal V}_N$ is given by suitable linear combinations of the elementary states \eqref{eq:elemstate2} with coefficients $b(y_1,\dots,y_N)$, which are complex-valued functions to be determined:
\begin{equation}\label{eq:psiN}
\Psi_N = \sum_{1 \le y_1 \le \dots \le y_N \le L} b(y_1,\dots,y_N) \vert y_1,\dots,y_N \rangle,
\end{equation}
and we assume again a plane wave decomposition for the functions $b(y_1,\dots,y_N)$ (the notations are similar to the ones used for the CBA based on $|\Omega\rangle$):
\begin{equation}
b(y_1,\dots,y_N) = \sum_{\sigma \in {\mathfrak S}_N} B_\sigma^{(j_1,\dots,j_P)}(k_1,\dots,k_N) \exp\left(\sum_{n=1}^N ik_{\sigma(n)} y_n \right) = \sum_{\sigma \in {\mathfrak S}_N} B_\sigma^{(j_1,\dots,j_P)}(\vec{k}) e^{i\vec{k_\sigma}\cdot\vec{y}}.
\label{eq:planewave2}
\end{equation}
The energy of the eigenstate $\Psi_N$ is then given by
\begin{equation}
\widetilde E_N = Lv_{22} + N(v_{21}+v_{12}-2v_{22}) + \sum_{n=1}^M (t_3\,e^{ik_{n}}+s_3\,e^{-ik_{n}})
\label{eq:energie2}
\end{equation}
By consistency, one has of course $\{ \widetilde E_{2L-M} \} = \{ E_M \}$ for a given value of $M$. \\
The Bethe equations has the same shape
\begin{equation}
e^{ik_{j}L} = \prod_{n \ne j} \widetilde S(k_n,k_j)\,, \qquad j=1,...,M.
\end{equation}
where now the $S$-matrix is the image of the original one by the transformation \eqref{eq:transfoC}.

\pagebreak[4]

\end{document}